\documentclass[prl,showpacs]{revtex4}
\usepackage{graphicx}
\usepackage{color}
\newcommand{\om}{\omega}

\newcommand{\E}{{\bf E}}
\newcommand{\ev}{{\bf e}}

\newcommand{\kv}{{\bf k}}
\newcommand{\K}{{\bf K}}
\newcommand{\rv}{{\bf r}}
\newcommand{\R}{{\bf R}}

\newcommand{\Hv}{{\bf H}}

\newcommand{\Sv}{{\bf S}}

\newcommand{\uv}{{\bf u}}

\newcommand{\shat}{\hat{s}}

\newcommand{\pup}{\hat{p}_1^+}

\newcommand{\pdp}{\hat{p}_2^+}

\newcommand{\mv}{{\bf m}}
\newcommand{\dv}{{\bf d}}

\begin{document}

\title{Thermal Emission by a Subwavelength Aperture}
\author{Karl Joulain,Youn\`es Ezzahri}
\affiliation{Institut Pprime, Universit\'e de Poitiers-CNRS-ENSMA, 86000 Poitiers, France}
\author{R\'emi Carminati}
\affiliation{ESPCI ParisTech, PSL Research University, CNRS, Institut Langevin, 1 rue Jussieu, F-75005, Paris, France}
\date{\today}
\begin{abstract}
We calculate, by means of fluctuational electrodynamics, the thermal emission of an aperture separating from the outside, vacuum or a material at temperature $T$. We show that thermal emission is very different whether the aperture size is large or small compared to the thermal wavelength. Subwavelength apertures separating vacuum from the outside have their thermal emission strongly decreased compared to classical blackbodies which have an aperture much larger than the wavelength. A  simple expression of their emissivity can be calculated and their total emissive power scales as $T^8$ instead of $T^4$ for large apertures. Thermal emission of disk of materials with a size comparable to the wavelength is also discussed. It is shown in particular that emissivity of such a disk is increased when the material can support surface waves such as phonon polaritons.
\end{abstract}
\pacs{ 44.40.+a,05.40-a,78.67.Pt}

\maketitle

\section{Introduction}
Since the end of the 19th century and the work of Max Planck, it has been known that thermal emission of radiation follows universal laws. For instance, the emissive power of a body at temperature $T$ cannot exceed the value given by the so-called Stefan law, that reads as $H^0(T)=\sigma T^4$, with $\sigma=5.67$ 10$^{-8}$ W.m$^{-2}$.K$^{-4}$. Another feature is that the thermal emission spectrum is broadband and peaked around $\lambda_m$ (given by the Wien law $\lambda_m T=2898 \, \mu$m.K), with a typical bandwidth of a few $\lambda_m$. However, theoretical models based on a fluctuationnal electrodynamics formalism have shown that thermal emission could deviate from the above mentioned behaviors when the length scales involved are small compared to the typical wavelength $\lambda_m$ of the emitted radiation. For example, when two heated bodies are separated by a small gap, radiative heat transfer surpasses that predicted by classical formulas, due to the coupling of evanescent modes on the surface of each body \cite{Polder:1971uu,Loomis:1994tx}. Heat transfer is enhanced in this case, and can even be dominated by transfer through modes at specific frequencies, especially when the materials exhibit resonances such as surface phonon or surface plasmon polaritons \cite{Shchegrov:2000td,Mulet:2002we,Joulain:2003hc,Joulain:2005ih}. Moreover, micro or nanostructured surfaces, such as periodic gratings, can scatter the thermaly excited evanescent waves into the far field, which substantially changes the emission properties. This mechanism has paved the way towards the design and fabrication of coherent thermal sources exhibiting both temporal and spatial coherence \cite{Greffet:2002ur}. Another way to couple the near field and the far field is to use the tip of a Scanning Near-Field Optical Microscopy and bring it at a submicron distance from the heated surface. The thermally populated evanescent modes can be coupled to a detector in the far field by scattering at the tip. This process underlies the principle of Thermal Radiation Scanning Tunneling Microscopy \cite{DeWilde:2006kta,Babuty:2013eb,Joulain:2014eu} (TRSTM), an imaging technique among others \cite{Jones:2012fx} that uses thermal radiation to perform imaging and spectroscopy of subwavelength structures.

The purpose of this paper is to explore another aspect of thermal emission at subwavelength scale. We study the conceptually simple situation of thermal emission by an aperture. Note that the problem could be addressed using the reciprocity theorem of electromagnetism. Indeed, in the theory of thermal radiation, it is known that reciprocity is the foundation of  Kirchhoff's law, stating that the emissivity of a material equals its absorptivity. This means that knowing the absorption efficiency $Q_{abs}(\omega)$ of a body at a given temperature $T$, the thermally emitted flux by this body at the same temperature and at frequency $\omega$ is given by \cite{Bohren:1983wi}
\begin{equation}
\label{emitq}
\phi(\omega,T)=Q_{abs}(\omega)\frac{\hbar\omega^2}{4\pi^2c^2[\exp(\hbar\omega/k_bT)-1]}
\end{equation}
where $\hbar$ is the reduced Planck constant and $k_b$ is Boltzmann's constant.
Therefore, the knowledge of the light absorption properties of an object at a given frequency allows one to deduce its thermal emission properties. For example, a sphere of a homogeneous material will emit according to Eq.~(\ref{emitq}) with $Q_{abs}$ given by the Mie theory \cite{vandeHulst:1981tc}.

In this paper, we address the problem from a different point of view. We use fluctuationnal electrodynamics in order to compute directly the thermal emission by an aperture. The principle of the approach is the following. In a body at local thermal equilibrium, temperature initiates fluctuating currents that radiates an electromagnetic field \cite{Rytov:1989ur}. Thermal currents are characterised statistically by a correlation function given by the fluctuation-dissipation theorem. Radiation by these currents is calculated by solving Maxwell's equations in the specific geometry, as in a standard antenna radiation problem. Note that in the specific case of an aperture, the emitted heat flux is given by the flux of the Poynting vector through a plane parallel to the aperture, allowing us to connect this flux to the Wigner transform of the electric field spatial correlation function \cite{WALTHER:1968cc,Wolf:1978cf,Apresyan:1996vq}. These spatial correlations are directly computed in fluctuationnal electrodynamics \cite{Henkel:2000tr,Carminati:1999uh}. We first focus on the simple case of an aperture separating vacuum at thermal equilibrium from the outside. Then the formalism is also applied to the  case of an aperture separating a material supporting resonant surface waves at thermal equilibrium from the outside

\section{Emissivity of an aperture}

\begin{figure}
\begin{center}
\includegraphics[width=7cm]{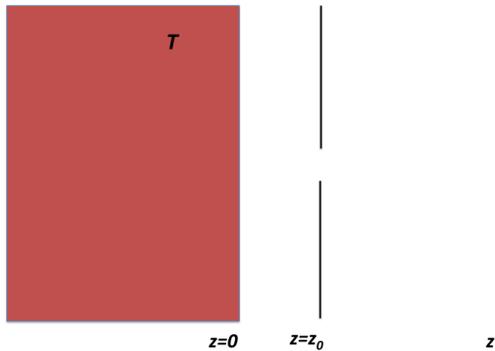}
\caption{Geometry of the model system. From the electromagnetic field in the plane $z=z_0$, one deduces the field and the radiated power
in any plane at a distance $z$.}
\label{System}
\end{center}
\end{figure}

The system considered here is depicted in Fig. \ref{System}. A semi-infinite material at temperature $T$ fills the half-space $z<0$, on top of which a mask with a transmission function $\tau(\R)$ is placed in a plane $z=z_0$, where $z_0\to 0$. Thermal radiation is emitted by the material through the mask, and the 
radiated power is calculated in a plane at a distance $z$ through the evaluation of the flux of the Poynting vector across this plane.

For monochromatic fields, the complex amplitude of the electric field $\E(\rv)$ in the plane $z$ can be written as a plane-wave expansion in the form
\begin{equation}
\label{ }
\E(\rv)=\int \E(\K,z_0)e^{i\K.\R}e^{i\gamma(z-z_0)}\frac{d^2\K}{4\pi^2}
\end{equation}
where $\kv=(\K,\gamma)=K\uv_\perp+\gamma \ev_z$ , $\rv=(\R,z)$ and $\gamma^2+\K^2=k_0^2$ with $k_0 = \om/c=2\pi/\lambda$.
The amplitude $\E(\K,z_0)$ of the plane waves in this expansion is the Fourier transform of the field in the plane $z=z_0$, and reads as
\begin{equation}
\label{EKR}
\E(\K,z_0)=\int \E(\R_0,z_0)e^{-i\K.\R_0}d^2\R_0 \ .
\end{equation}
The power $\phi(\om)$ radiated in the far field is defined as the flux of the Poynting vector through the plane $z$. For monochromatic fields, the time-averaged Poynting  vector is $\Sv(\rv)=1/2\Re\left[\E(\rv)\times\Hv^*(\rv)\right]$, where $\Hv(\rv)$ is the complex amplitude of the magnetic field and the superscript $*$ stands for complex conjugate. Using the Maxwell equation  $\nabla \times \E = i\omega\mu_0 \Hv$ and the plane-wave expansion of the electric field, one obtains 

\begin{equation}
\label{phiK}
\phi(\om)= \frac{1}{2\mu_0\om}\Re\int\gamma|\E(\K,z_0)|^2\frac{d^2\K}{4\pi^2} \ .
\end{equation}
Note that the integration is restricted to propagating waves, {\it i.e}, waves for which $K<k_0$ since $\Re(\gamma)=0$ when $K>k_0$. In this case, this integration can also be understood as an angular integration on the upper hemisphere of the wavevector $\kv$ with constant modulus $|\kv|=\om/c$. 

Equation (\ref{phiK}) shows that the knowledge of the field in the plane $z=z_0$ permits an explicit calculation of the radiative flux emitted in the far field. In our model, this field can be understood as the field radiated by the semi-infinite medium and transmitted through the aperture. Denoting by $\E^{inc}$ the field right before the plane of the aperture, and describing the aperture (or actually any scattering object placed in the plane $z=z_0$) by a transmission matrix $\tau_{ij}(\K,\K')$, one can write the field in the plane $z=z_0$ as
\begin{equation}
\label{}
E_i(\K,z_0)=\int \tau_{ij}(\K,\K')E_j^{inc}(\K',z_0)\frac{d^2\K'}{4\pi^2} \ .
\end{equation}
Inserting this expression into Eq. (\ref{phiK}) leads to
\begin{equation}
\label{eq:phi1}
\phi(\omega)  =  \frac{1}{32\mu_0\omega\pi^6}\int\gamma \, \tau_{ij}(\K,\K') \tau_{ik}^*(\K,\K'')E_j^{inc}(\K',z_0)E_k^{inc*}(\K'',z_0)d^2\K d^2\K'd^2\K''
\end{equation}

The incident field can be calculated as the field radiated by the semi-infinite material in absence of the aperture (this is the simplest model, a self-consistent calculation being outside the scope of the present study). This field is linearly related to the thermally excited electric currents inside the material, through a relationship of the form
\begin{equation}
\label{ }
E_i^{inc}(\rv)=i\mu_0\omega\int d^3\rv' G_{im}(\rv,\rv')j_m(\rv') 
\end{equation} 
where $\mathbf{G}$ is the tensor Green function that describes the electrodynamic response of the semi-infinite material and $\mathbf{j}$ is the electric current density. The Green function in this geometry can be written as a plane-wave expansion that involves the Fresnel transmission factors at the interface $z=0$ between the medium and vacuum \cite{Sipe:1987td}.  According to this expansion, the incident electric field reads
\begin{eqnarray}
E_i^{inc}(\rv) & = &\frac{ -\mu_0\omega}{8\pi^2} \int \frac{d^3\rv' d^2\K}{\gamma_2}\left [\ev_i \shat t_{21}^s \shat+\pup t_{21}^p\pdp\ev_m    \right ] e^{i\K.(\R-\R')}e^{i\gamma z}e^{-i\gamma_2 z'}j_m(\rv')\\
 & = & \int E_i^{inc}(\K,z_0)\frac{d^2\K}{4\pi^2} 
\end{eqnarray}
where $\shat=\K/|K|\times \ev_z$, $\hat{p}_i^{+}=\left[K^2\ev_z-\gamma_i K_x \ev_x-\gamma_iK_y\ev_y\right]/(n_ik_0K)$, and $t_{21}^s$ and $t_{21}^p$ are the Fresnel transmission factors for $s$ and $p$ polarization, respectively \cite{Sipe:1987td}.
By identification, one obtains the expression of the Fourier transform of the incident field in the plane $z=z_0$~:
\begin{equation}
\label{eq:Eincj}
E_i^{inc}(\K,z_0)=\frac{-\mu_0\omega}{2}\int\frac{d^3\rv'}{\gamma_2}\left [\ev_i \shat t_{21}^s \shat+\pup t_{21}^p\pdp\ev_m    \right ] e^{-i\K.\R'}e^{i\gamma_1z_0}e^{-i\gamma_2z'}j_m(\rv') \ .
\end{equation}
The thermally excited currents are fluctuating fields, that are describes statistically. In order to compute fluxes, one needs second order quantities. The spatial correlation function of the currents in the material at thermal equilibrium is given by the fluctuation-dissipation theorem
\begin{equation}
\label{eq:FDTj}
\langle j_k(\rv,\omega)j_l(\rv',\omega')\rangle=\frac{\epsilon_0\Im[\epsilon(\omega)] \, \omega \, \Theta(\omega,T)}{\pi}\delta_{kl}\delta(\rv-\rv')\delta(\omega-\omega')
\end{equation}
where the brackets denote an average over thermal fluctuations, $\Theta(\omega,T)=\hbar\omega/[\exp(\hbar\omega/k_bT)-1]$, and $\epsilon(\omega)$ is the dielectric function of the medium.
From Eqs. (\ref{eq:phi1}), (\ref{eq:Eincj}) and (\ref{eq:FDTj}), one obtains the following expression of the thermally radiated flux 
\begin{equation}
\label{phitot}
\phi(\omega,T)=\frac{\Theta(\omega,T)}{32\pi^5}\int d^2\K d^2\K' \gamma(\K)\frac{\Re(\gamma_2(\K'))}{|\gamma_2(\K')|^2}e^{-2\Im(\gamma_1(\K))z_0}\tau_{ij}(\K,\K') \tau_{ik}^*(\K,\K')\left(M^s_{jk}(\K')+M^p_{jk}(\K')\right)
\end{equation}
where
\begin{equation}
\label{ }
M^s(\K)=\frac{|t_{21}^s|^2}{K^2}\left(\begin{array}{ccc}
 K_x^2     & -K_xK_y & 0 \\
 -K_xK_y     & K_y^2  & 0\\
     0 & 0 &0
\end{array}\right)
\end{equation}
and
\begin{equation}
\label{ }
M^p(\K)=|t_{21}^p|^2\frac{|\gamma_2|^2+K^2}{|n_2|^2|n_1|^2k_0^4K^2}\left(\begin{array}{ccc}
 |\gamma|^2K_x^2     &   |\gamma|^2K_xK_y   & -\gamma K_xK^2   \\
   |\gamma|^2K_xK_y    & |\gamma|^2K_y^2   &  -\gamma K_yK^2  \\
  -\gamma^* K_xK^2    &  -\gamma^* K_yK^2  & K^4
\end{array}\right) \ .
\end{equation}
Note that this final expression is restricted to positive frequencies only (as is usual in radiative transfer), which implicitely assumes that all fields in the derivation have been replaced by their analytic signals (in practice this results in an extra factor of 4, see \cite{Joulain:2005ih} for details). 

This expression of the radiated power appeals for the definition of an effective emissivitty. Indeed, in the framework of geometrical optics, 
the emitted flux by an object with surface $S$ is usually written in the form 
\begin{equation}
\phi(\omega,T)=\varepsilon\frac{\Theta(\omega,T) \, \omega^2}{4\pi^2 c^2}S
\end{equation}
where $\varepsilon$ is by definition the emissivity of the object. From Eq.~(\ref{phitot}) one can define the {\it effective} emissivity of the aperture [or of any scattering object defined by a transmission matrix $\tau_{ij}(\K,\K')$] as
\begin{equation}
\label{ }
\varepsilon_{eff}=\frac{1}{8\pi^3k_0^2S}\int  d^2\K d^2\K'  \gamma(\K)\frac{\Re[\gamma_2(\K')]}{|\gamma_2(\K')|^2}e^{-2\Im[\gamma_1(\K)]z_0}\tau_{ij}(\K,\K') \tau_{ik}^*(\K,\K')\left [M^s_{jk}(\K')+M^p_{jk}(\K')\right ]
\end{equation}
This is the general expression of the emissivity of an aperture defined by its transmission matrix $\tau_{ij}(\K,\K')$. It involves a double integral the transmission matrix over all parallel wavevector. Integration over $\mathbf{K}$ is limited to propagative waves such as $K \leq k_0$, whereas integration over $\mathbf{K}'$ includes {\it a priori} both propagating ($K' \leq k_0$)  and evanescent waves ($K' > k_0$). The contribution of evanescent waves to the radiated flux in the far field results from a scattering process. The thermally excited evanescent waves with large wavevectors $K' > k_0$ are scattered into propagating waves with $K \leq k_0$ by scattering at the aperture. Another feature of the expression of the effective emissivity is that the material and geometrical resonances are contained in the integral both in the transmission matrix $\tau_{ij}(\K,\K')$ and the Fresnel transmission factors. Finally, note that due to reciprocity, the expression of the emissivity can also be seen as that of the absorption cross-section normalized by the geometrical cross-section $S$.

\section{Aperture in vacuum}

As the simplest example, we consider the case of blackbody radiation in a vacuum at temperature $T$ transmitted through an aperture in an opaque screen. In the general model derived in the preceding section, this amounts to considering a material with transmission factors $t^s$ and $t^p$ equal to unity. The radiative heat flux coming out from the aperture can be calculated analytically in two asymptotic cases. The first case corresponds to an aperture with a radius $r_0$ much larger than the typical thermal wavelength. Under this assumption, one can make use of the Kirchhoff approximation in which the field equals the incident field in the aperture and vanishes outside. The limit of validity of the Kirchhoff approximation is estimated to be $k_0r_0\sim 6$, which corresponds to an aperture radius on the order of the wavelength \cite{Levine:1950td}. Under this assumption, the transmission matrix is reduced to a scalar so that $\tau_{ij}(\K,\K')=\delta_{ij}T(\K-\K')$, where
 \begin{equation}
\label{ }
T(\K)=\int T(\R) e^{-i\K.\R}d^2\R
\end{equation}
and $T(\R)=1$ inside a circle of radius $r_0$ (the aperture) and $T(\R)=0$ outside. An explicit calculation leads to
\begin{equation}
\label{ }
T(\K)=\int_0^{2\pi}d\varphi\int_0^{r_0}Re^{-iKR\cos\varphi}dR=2\pi\int_0^{r_0}RJ_0(KR)dR=\pi r_0^2\left(\frac{2J_1(r_0 K)}{r_0 K}\right) \ .
\end{equation}
Inserting this expression of the transmission matrix into Eq.~(\ref{phitot}) allows in principle to calculate the radiated flux. It is however easier to rewrite the flux as
\begin{equation}
\label{ }
\phi(\omega,T)=\frac{\Theta(\omega,T)}{16\pi^5}\int\frac{\gamma(\K)}{\gamma(\K')}T(\R')T(\R'')e^{-i(\K-\K').\R'}e^{i(\K-\K').\R''}d^2\K d^2\K'd^2\R'd^2\R''
\end{equation}
and to perform the change of variables $\mv=(\R'+\R'')/2$ and $\dv=\R'-\R''$, leading to
\begin{equation}
\label{ }
\phi(\omega,T)=\frac{\Theta(\omega,T)}{16\pi^5}\int\frac{\gamma(\K)}{\gamma(\K')}T(\mv+\dv/2)T(\mv-\dv/2)e^{-i(\K-\K').\dv}d^2\K d^2\K'd^2\mv d^2\dv \ .
\end{equation}
Since the product $T(\mv+\dv/2)T(\mv-\dv/2)$ is independent on the variable $\mv$, the integration over $\mv$ gives
\begin{equation}
\label{ }
\int T(\mv+\dv/2)T(\mv-\dv/2) d\mv=\pi r_0^2 W(d)=\pi r_0^2\times \frac{2}{\pi}\left[\arccos\frac{d}{2r_0}-\frac{d}{2r_0}\sqrt{1-\left(\frac{d}{2r_0}\right)^2}\right] \ .
\end{equation}
Using spherical coordinates with angles $\theta$ and $\varphi$, one can write $\kv=(\K,\gamma)=k_0(\sin\theta\cos\varphi,\sin\theta\sin\varphi,\cos\theta)$
and transform the integral into
\begin{equation}
\label{ }
\phi(\omega,T)=\frac{\Theta(\omega,T)}{16\pi^5}\pi r_0^2 k_0^4\int\cos^2\theta\sin\theta e^{-ik_0 d\sin\theta\cos\varphi}\sin\theta'  e^{ik_0 d\sin\theta'\cos\varphi'}W(d)d\theta d\varphi d\theta'd\varphi' d^2\mathbf{d}
\end{equation}
which, after integration over azimuthal angles, gives
\begin{equation}
\label{ }
\phi(\omega,T)=\frac{\Theta(\omega,T)}{16\pi^5}\pi r_0^2 k_0^44\pi^2\int\cos^2\theta\sin\theta J_0(k_0d\sin\theta)\sin\theta' J_0(k_0d\sin\theta') W(d)d\theta d\theta'd^2\mathbf{d} \ .
\end{equation}
Integration over $\theta$ and $\theta'$, knowing that $\mathbf{d}$ extends over a disk of radius $2r_0$, leads to
\begin{eqnarray}
\label{ }
\phi(\omega,T)&=&\frac{\Theta(\omega,T)k_0^2}{4\pi^2}\pi r_0^2 2\int_0^{2k_0r_0}W(u/k_0)\sin u F(u) du \nonumber \\
&=&\phi^0(\om)\int_0^{2k_0r_0}W(u/k_0)2\sin u F(u) du\nonumber\\
&=&\phi^0(\om)\varepsilon_{vac}^{eff}(\omega)
\end{eqnarray}
where $F(u)=(\sin u-u\cos u)/u^3$. The last expression defines the effective emissivity $\varepsilon_{vac}^{eff}(\omega)$ at frequency $\omega$ of a blackbody of circular radius $r_0$. 

When the aperture is large compared to the wavelength, thermal emission corresponds to a blackbody. However, our result shows that the emissivity of an aperture is smaller than 1 if the aperture size is on the order of the wavelength. Pushing the Kirchhoff approximation at its limit $k_0r_0=6$, we obtain $\varepsilon_{vac}^{eff}\simeq 0.84$. This can be easily understood since waves with wavelengths on the order or smaller than the aperture size can hardly be transmitted. The aperture acts as a high pass filter, reducing the contribution of low frequency waves, which is a feature of the underlying diffraction process.
However, it is known that the Kirchhoff approximation breaks down when the aperture size becomes smaller than the wavelength \cite{Bethe:1944va,Levine:1950td,Bouwkamp:1954wt}, typically when $k_0r_0<6$. Bethe \cite{Bethe:1944va} and Bouwkamp \cite{Bouwkamp:1954wt} have indeed shown that the transmission through a small hole is actually weaker than that predicted by the Kirchhoff approximation. The problem addressed by Bethe and Bouwkamp's theory is that of transmission through a hole in a perfectly conducting screen. By introducing fictitious magnetic charges and currents in the diffracting hole satisfying boundary conditions on the screen, their theory allows one to calculate the scattering cross-section and the transmission matrix $\tau(\K,\K')$ in the regime $k_0r_0\ll 1$.   
One ends up with
\begin{equation}
\label{ }
|\tau_{ik}^s(\K,\K')|^2=\frac{64}{9}k_0^2r_0^6\frac{\cos^2\theta'}{\cos^2\theta}(1-\sin^2\theta\cos^2\varphi)
\end{equation}
for $s$ polarization, and with
\begin{equation}
\label{ }
|\tau_{ik}^p(\K,\K')|^2=\frac{64}{9}k_0^2r_0^6\frac{\cos^2\theta+\sin^2\theta(\cos^2\varphi+1/4\cos^2\varphi')-\sin\theta\cos\varphi\sin\theta'}{\cos^2\theta}
\end{equation}
for $p$ polarization.
Let us note that the transmission matrix is here limited to propagative waves ($K,K'\leq k_0$). Inserting these two expressions into Eq. \ref{phitot}, one can perform the integration over incoming and outgoing wavevectors ($\K$ an $\K'$), which for propagating waves amounts to integrating over $\theta$, $\varphi$, $\theta'$ and $\varphi'$. This leads to the following expression of the radiative thermal flux emitted by a subwavelength hole~:
\begin{equation}
\label{eq:flux_small_hole}
\phi(\omega,T)=\frac{16}{27}\frac{k_0^6r_0^6}{\pi^3}\Theta(\omega,T)=\frac{64k_0^4r_0^4}{27\pi^2}\phi^0(\omega,T)=\varepsilon_{eff}\phi^0(\omega,T) \ .
\end{equation}
This result shows that the effective emissivity of a subwavelength hole is $\varepsilon_{eff}=64(k_0r_0)^4/(27\pi^2)$. As expected, this emissivity is smaller than that predicted by the Kirchhoff approximation, which predicts a scaling in $k_0^2r_0^2$. Note that the scaling in $k_0^4r_0^4$ that is obtained for a subwavelength hole is consistent with that expected for Rayleigh scattering ({\it i.e.} scattering by particles much smaller than the wavelength). This result confirmes that small apertures behave as high-pass filters regarding thermal emission. 

Expression (\ref{eq:flux_small_hole}) gives the radiative flux at a given frequency $\omega$. If the condition $k_0r_0\ll 1$ is satisfied on the full spectral range covered by thermal emission (typically $\lambda_m/2<\lambda<5\lambda_m$ in terms of wavelengths), the spectrally integrated flux can be calculated, and reads
\begin{equation}
\label{ }
\phi=\int_0^\infty\frac{16}{27}\frac{k_0^6r_0^6}{\pi^3}\Theta(\omega,T)d\omega=\frac{128 r_0^4\pi^4k_b^8T^8}{405c^6\hbar^7}\times\pi r_0^2 \ .
\end{equation}
It is interesting to note that instead of following the usual $T^4$ law of free-space blackbody radiation, the power emitted by a subwavelength blackbody follows a $T^8$ law. This means that for a given aperture size $r_0$, when the temperature is decreased so that $\lambda_m$ is larger than $r_0$, the thermally emitted power decreases drastically, much faster than predicted by the usual Stefan-Bolztmann law. For example, a hole with $r_0=1\mu$m at 77 K (liquid Nitrogen temperature) has an emissive power of 
1.99 W.m$^{-2}$ according to Stefan-Boltzmann law, and of $4.75\times$ 10$^{-4}$ W.m$^{-2}$ according to the law derived in this paper using the Bethe-Bouwkamp theory.
Finally, let us remark that deriving an analytical expression of the emissivity in the intermediate regime $k_0r_0\sim 1$ is out of reach. In that case, one should follow approaches that have been used, for example, to address the problem of extraordinary transmission through subwavelength holes \cite{Nikitin:2008ga,GarciaVidal:2010ed,Nikitin:2010fr} and compute the absorption efficiency, that directly leads to the emissivity according to Kirchhoff's law.

\section{Aperture filled with a material}

In this section we address the thermal emission by an aperture when the medium occupying the half-space $z<0$ is a real material (see the geometry in Fig.~1). This problem cannot be solved in its full generality since there is no exact expression of the transmission matrix $\tau$ valid for any material. However, the Kirchhoff approximation can be used as long as $k_0r_0\simeq 6$, and we limit the study to that regime. This will allows us to highlight interesting phenomena that occur when the aperture size approaches the wavelength. Under the Kirchhoff approximation, the emitted radiative flux reads
\begin{eqnarray}
\label{formflux}
\phi(\omega)&=&\phi^0(\omega)\int_0^{2k_0r_0}W(u/k_0)u F(u) du \nonumber\\
&\times&\left\{\int_0^{1}\frac{\kappa J_0(\kappa u) d\kappa}{\sqrt{1-\kappa^2}}(2-|r^s|^2-|r^p|^2)
+\int_{1}^\infty\frac{2\kappa J_0(\kappa u) d\kappa}{\sqrt{\kappa^2-1}}\left[\Im(r^s)+(2\kappa^2-1)\Im(r^p)\right]e^{-2\sqrt{\kappa^2-1}k_0z}\right\}
\end{eqnarray}
where $\kappa=K/k_0$. This expression contains two contributions: the propagating wave contribution for $\kappa<1$ and the evanescent wave contribution for $\kappa>1$. Let us first check that from Eq.~(\ref{formflux}) one recovers the classical expression of the radiative flux when the aperture size is much larger than the wavelength. For a circular aperture, this corresponds to the condition $k_0r_0\gg1$. Let us note that $W(u/k_0)$ decreases smoothly from 1 to 0 when $u/k_0$ varies from 0 to $2r_0$. $F(u)$ decreases fastly to 0 when $u$ is large compared to 1. When $k_0r_0\gg1$, there is a domain in which $u\gg 1$ and $u\ll 2k_0r_0$. In this domain, the upper bound of integration over $u$ in  Eq.~(\ref{formflux}) can be replaced by $\infty$, and $W(u/k_0)$ can be replaced by 1. Noting that $\int_0^\infty uF(u)J_0(\kappa u)du$ vanishes if $\kappa>1$ and $\sqrt{1-\kappa^2}$ if $\kappa<1$ \cite{Gradshteyn:2007uy}, one retrieves that there is no contribution of the evanescent waves to the emitted flux for large apertures. Moreover, the expression of the emitted flux equal the classical expression
\begin{equation}
\label{ }
\phi=\phi_{clas}=\phi^0(\om)\int_0^1\kappa d\kappa (2-|r^s|^2-|r^p|^2)
\end{equation}
where the integral represents the emissivity of the material. Note that this emissivity is equal to 1 when the Fresnel reflection factors vanish, {\it i.e.} in the vacuum blackbody radiation limit.

\begin{figure}
\begin{center}
\includegraphics[width=10cm]{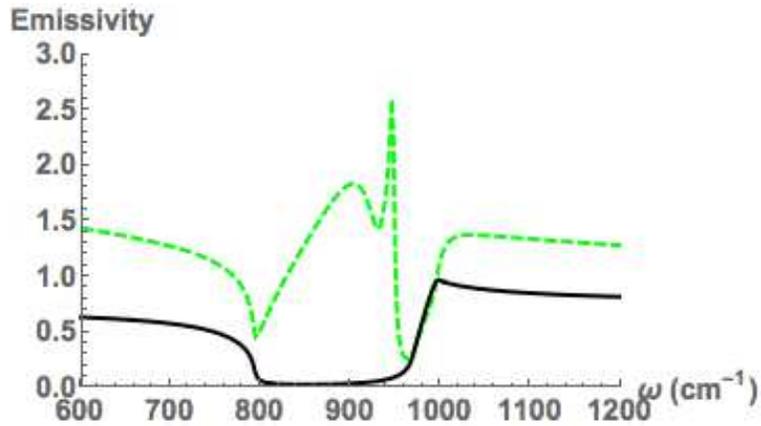}
\caption{Emissivity vs angular frequency for a circular aperture filled of SiC with a radius of 100 $\mu$m (plain) and with a radius of 10 $\mu$m (Dashed).}
\label{EmisSiC}
\end{center}
\end{figure}

In the regime where $k_0r_0$ is not large compared to one, the contribution of the evanscent waves is no more negligible, and one has to integrate Eq.~(\ref{formflux}) numerically. An interesting situation is that of  a material supporting surface waves, such as SiC, at the limit of validity of the Kirchhoff approximation in terms of aperture size. In Fig.~\ref{EmisSiC}, the effective emissivity ({\it i.e.} $\phi/\phi^0$) is plotted versus frequency around the surface-phonon polariton resonance of SiC which occurs for $\lambda= 10.6 \, \mu$m. For an aperture with radius $r_0=100 \,\mu$m filled with SiC, the emissivity is the same as that obtained for a massive material. It is close to one in a broad specral range, except close to the surface-polariton resonance for which the material is very reflective. For a radius $r_0=10 \,\mu$m, the emissivity is enhanced in the spectral domain where SiC supports surface polaritons. These surface polaritons are thermally excited and scattered by the aperture, which adds new channels for far-field thermal radiation. One can even observe an effective emissivity larger than one around the surface polariton resonance frequency. This means that the thermal emission of the aperture is larger that the blackbody emissive power multiplied by the geometrical cross-section. A radiometric interpretation is that the effective aperture emission size is larger than it geometrical size. Using reciprocity (or Kirchhoff's law), one can also understand that the emissivity is equivalent to an absorption cross-section, normalized by the geometrical section. It is actually well-known in scattering theory that  scattering by nano-objects or nano-antennas such as nano-spheres or nano-cylinders leads to cross-section larger than the geometrical size. This is the so called antenna effect. Note however that when the surface considered for thermal emission becomes larger than the wavelength, there is no way that this emission can surpass blackbody emissive power. For example, it is not possible to make a macroscopic surface made of small aperture that overall would surpass blackbody limit. There is therefore no violation of the blackbody limit for macroscopic surface containing or not sub wavelength objects.

\section{Conclusion}
We have shown that thermal emission by a material can be substantially modified by confining this material to areas on the order or smaller that the typical emission wavelength. The confinement acts as a high pass filter, that changes the spectrum of thermal emission, as well as the value of the effective emissivity. In the case of a subwalength hole, the effective emissivity has been calculated using the Bethe-Bouwkamp model. It has been shown that in this limit, the emissivity scales as $k_0^4r_0^4$, and that total emitted flux scales as $T^8$, instead of the usual blackbody $T^4$ law. In the case of an aperture separating a material supporting surface modes (such as surface-phonon polaritons) from the outside, a contribution from evanescent waves scattered by the aperture generates an enhancement of the emissivity around the resonant frequency. From a thermal engineeering point of view, this study shows that the design of subwavelength scattering structures (the aperture being a simple example) could allow one to produce thermal sources with high spatial confinement and large efficiency at specific frequencies. The design of more complex structures would require an improvement of the theory to solve the full electrodynamic problem without requiring simple geometries or crude approximations. This could be done using numerical approaches already in use in nanophotonics, and in fluctuating electrodynamics such as discrete dipole approximation (DDA), finite-domain time difference (FDTD), or rigorous coupled wave algorithm (RCWA), to cite a few.

\section{Acknowledgements}

This work was supported by the French ANR through project number ANR-13-BS10-0013-04 (NATO project). This work also pertains to the French Government Program "Investissement d'avenir" (LABEX INTERACTIFS, ANR-11-LABX-0017-01)



\end{document}